\documentclass[a4paper]{jpconf}
\usepackage{graphicx}
\begin{document}
\title{VIP2 at Gran Sasso - Test of the validity of the spin statistics theorem for electrons with X-ray spectroscopy}

\author{J. Marton$^{1}$, A. Pichler$^{1}$, S. Bartalucci$^{2}$, M. Bazzi$^{2}$, S. Bertolucci$^{4}$, C. Berucci$^{1,2}$, M. Bragadireanu$^{2,3}$, M. Cargnelli$^{1}$, A. Clozza$^{2}$, C. Curceanu$^{2,3,9}$, L. De Paolis$^{2}$, S. Di Matteo$^{6}$, J.-P. Egger$^{7}$, C. Guaraldo$^{2}$, M. Iliescu$^{2}$, M. Laubenstein$^{8}$, E. Milotti$^{4}$, D. Pietreanu$^{2,3}$, K. Piscicchia$^{2,9}$, A. Scordo$^{2}$, H. Shi$^{2}$, D. Sirghi$^{2,3}$, F. Sirghi$^{2,3}$, L. Sperandio$^{2}$, O. Vazquez-Doce$^{2,10}$, E. Widmann$^{1}$ and J. Zmeskal$^{1}$}
\address{$^{1}$ Stefan Meyer Institute for subatomic physics, Boltzmanngasse 3, 1090 Vienna, Austria}
\address{$^{2}$ INFN, Laboratori Nazionali di Frascati, CP 13, Via E. Fermi 40, I-00044, Frascati (Roma),
Italy}
\address{$^{3}$ “Horia Hulubei”National Institute of Physics and Nuclear Engineering, Str. Atomistilor no.
407, P.O. Box MG-6, Bucharest - Magurele, Romania}
\address{$^{4}$ Dipartimento di Fisica, Universit{\`a}  di Trieste and INFN– Sezione di Trieste, Via Valerio, 2,
I-34127 Trieste, Italy}
\address{$^{5}$ University and INFN Bologna, Via Irnerio 46, I-40126, Bologna, Italy}
\address{$^{6}$ Institut de Physique UMR CNRS-UR1 6251, Universit{\'e} de Rennes1, F-35042 Rennes, France}
\address{$^{7}$ Institut de Physique, Universit{\'e} de Neuch\^{a}tel 1 rue A.-L. Breguet, CH-2000 Neuch\^{a}tel, Switzerland}
\address{$^{8}$ INFN, Laboratori Nazionali del Gran Sasso, S.S. 17/bis, I-67010 Assergi (AQ), Italy}
\address{$^{9}$ Museo Storico della Fisica e Centro Studi e Ricerche “Enrico Fermi”, Roma, Italy}
\address{$^{10}$ Excellence Cluster Universe, Technische Universit{\"ä}t  M{\"ü}nchen, Boltzmannstra{\ss}e 2, D-85748 Garching, Germany}

\ead{johann.marton@oeaw.ac.at}

\begin{abstract}
In the VIP2 (VIolation of the Pauli Exlusion Principle) experiment at the Gran Sasso underground laboratory (LNGS) we are searching for possible violations of standard quantum mechanics predictions. With high precision we investigate the Pauli Exclusion Principle and the collapse of the wave function  (collapse models). We will present our experimental method of searching for possible small violations of the Pauli Exclusion Principle for electrons, via the search for "anomalous" X-ray transitions in copper atoms, produced by "new" electrons (brought inside a copper bar by circulating current) which could have the probability to undergo Pauli-forbidden transition to the ground state (1 s level) already occupied by two electrons. We will describe the concept of the VIP2 experiment taking data at LNGS presently. The goal of VIP2 is to test the PEP for electrons with unprecedented accuracy, down to a limit in the probability that PEP is violated at the level of 10$^{-31}$. We will show preliminary experimental results obtained at LNGS and discuss implications of a possible violation.
\end{abstract}

\section{Introduction}
Wolfgang Pauli discovered the Exclusion Principle (PEP) named after him which could explain the periodic table of the elements \cite{Pauli1925}. Among the known rules of nature PEP  is an outstanding one, which can explain important phenomena like the stability of matter, the existence/stability of neutron stars and many others. Nowadays we can trace back the PEP to the spin-statistics theorem which classifies nature according to the spin in in fermionic (odd spin)  and bosonic (even spin)  systems. Remarkably no simple intuitive explanation for the PEP could be given. Several proofs of the PEP are based on complicated arguments can be found in the literature \cite{Pauli1940,Luders1958}.
The proof by L\"uders and Zumino \cite{Luders1958} is based on a clear set of assumptions:

\begin{itemize}
  \item[-] Invariance with respect to the proper inhomogeneous Lorentz group
  \item[-] Two operators of the same field at points separated by a spacelike interval either commute or anticommute (locality)
  \item[-] The vacuum is the state of lowest energy
  \item[-] The metric of the Hilbert space is positive definite
  \item[-] The vacuum is not identically annihilated by a field
\end{itemize}

If at least one of these assumptions is invalid then a violation of the Pauli Principle would be possible.
There are also theoretical attempts to accomplish PEP violations. Some recent theoretical studies can be found in refs. \cite{Jackson2008a, Balachandran2010}.

\section{VIP2 Experiment}

\subsection{Method of PEP testing}

\begin{figure}[h]
  \centering
  \includegraphics[width=12cm]{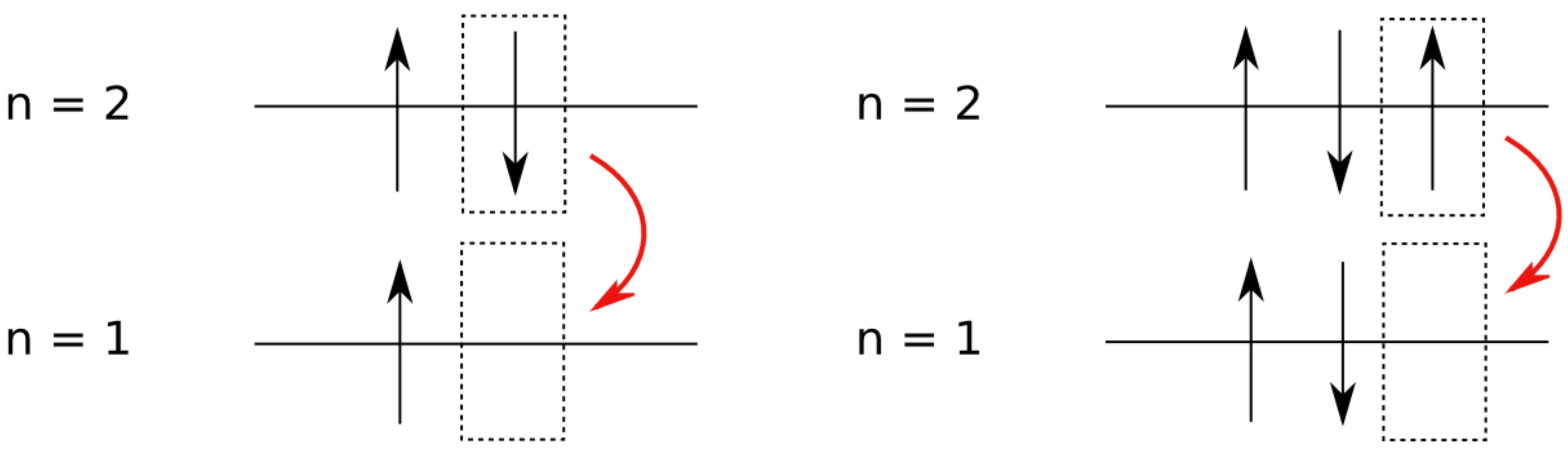}
  \caption{Transitions to the 1s ground state: Allowed transition 2p-1s (left) and Pauli forbidden transition to the fully occupied 1s state (right).}\label{scheme}
\end{figure}

The idea behind the VIP2 experiment follows an experiment performed by Ramberg and Snow \cite{Ramberg1990} with strongly improved signal sensitivity and background suppression. Like this experiment we search for Pauli forbidden X-ray transitions in copper after introducing "new" electrons to the system. The concept is based on the assumption that an electric current running through a copper conductor resembles a source of electrons which are "new" to the systems of copper atoms of the copper conductor. Thus one can search for Pauli-forbidden transitions in the copper atoms (see fig. \ref{scheme}). The transition energy of the PEP violating transition is shifted in energy due to the shielding by the "extra" electron in the 1s state. These shifted transition energies can be calculated using a multiconfiguration Dirac-Fock approach taking the relevant corrections (e.g. relativistic corrections) into account \cite{Sperandio2008, DiMatteo2005}.

\subsection{VIP2 Setup at LNGS}
An experiment VIP \cite{Collaboration2004, Bartalucci2006} following the concept of Ramberg and Snow was set up in the underground laboratory LNGS in Gran Sasso/Italy (LNGS). As X-ray detectors VIP used charge coupled devices (CCDs) \cite{Egger1993} providing  very good energy resolution, large detector solid angle and high intrinsic efficiency. The CCDs were already successfully employed in an experiment on kaonic atoms at LNF Frascati \cite{Beer2005, Ishiwatari2006a}. The CCDs ware positioned around a pure copper cylinder operated without and with up to 40 A current. The cosmic background in the LNGS site is strongly suppressed ($\sim 10^{-6}$) due to the rock coverage. Additionally the setup was covered by passive lead shielding.

\begin{figure}
  \centering
  \includegraphics[width=7cm]{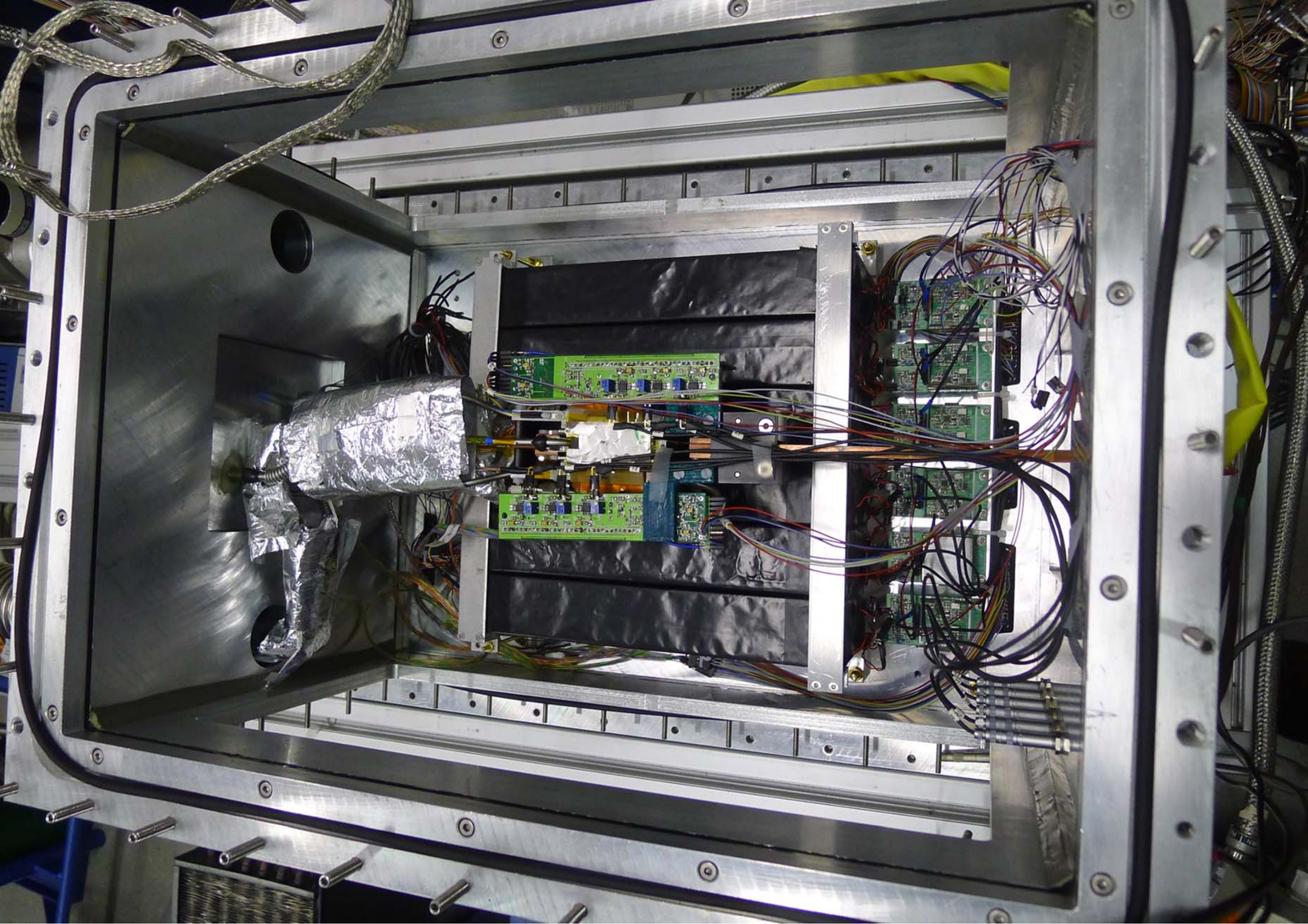}
  \caption{This photo shows the interior of the VIP2 experiment. In the box the copper target, the SDDs and the plastic scintillators are mounted. An insulation vacuum ($\sim$10$^{-5}$ mbar) inside the box is necessary to operate the SDDs at 100 K.}\label{vip2 box}
\end{figure}

To further enhance the sensitivity the experiment VIP2 with SDDs (Silicon Drift Detectors) as X-ray detectors was set up at LNGS. The experimental setup  provides a larger X-ray detector solid angle, higher current and is employing active shielding by plastic scintillators as background sensitive detectors. Due to the timing capability of SDDs the timing information of the SDD detectors and plastic scintillator signals can be used to additionally suppress background events.

\begin{figure}[h]
  \centering
  \includegraphics[width=10cm]{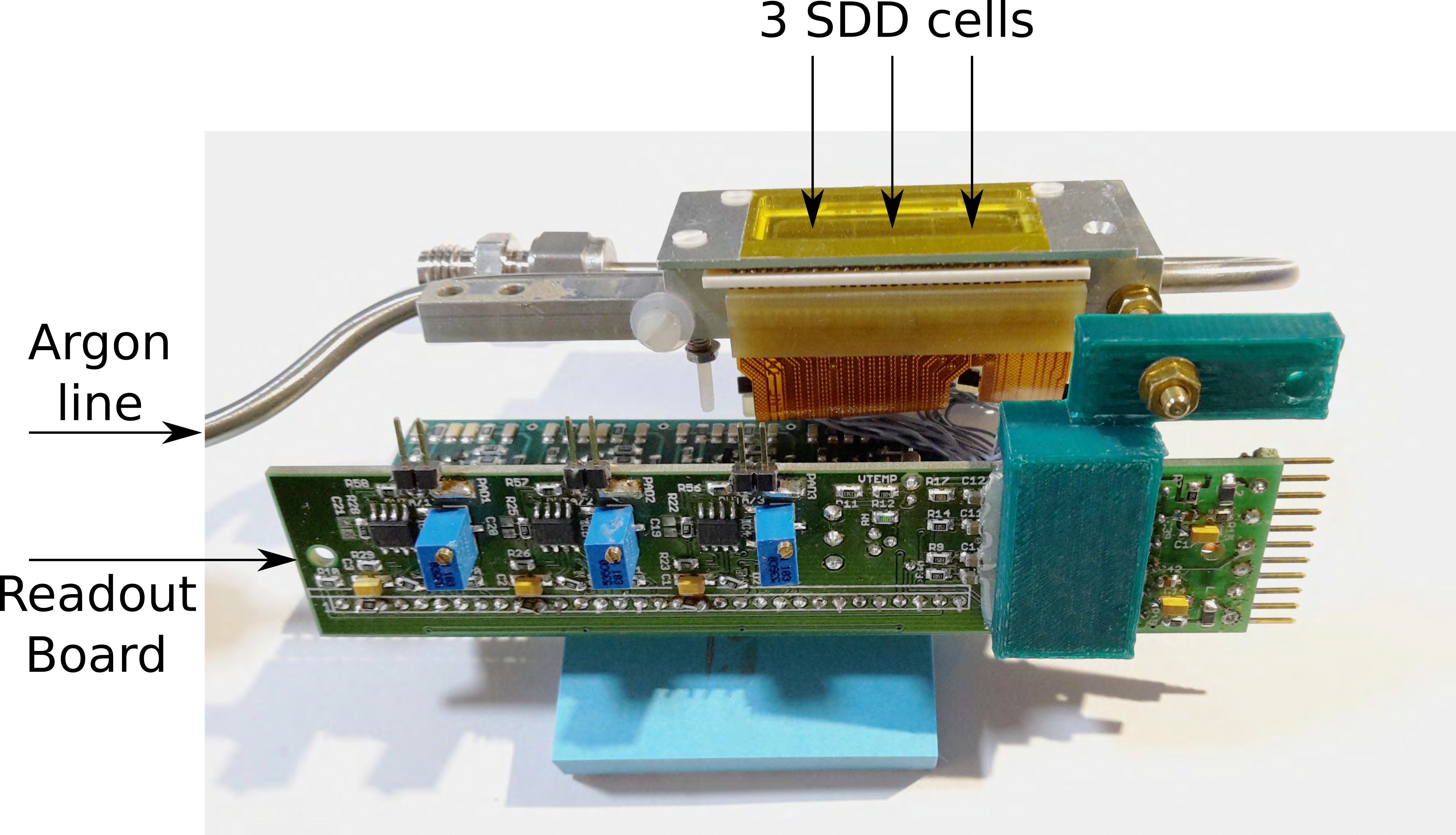}
  \caption{The VIP2 x-ray detector with 3 SDDs mounted.}\label{vip2-xray}
\end{figure}

\section{Preliminary Results} 
\begin{figure}
  \centering
  \includegraphics[width=9cm]{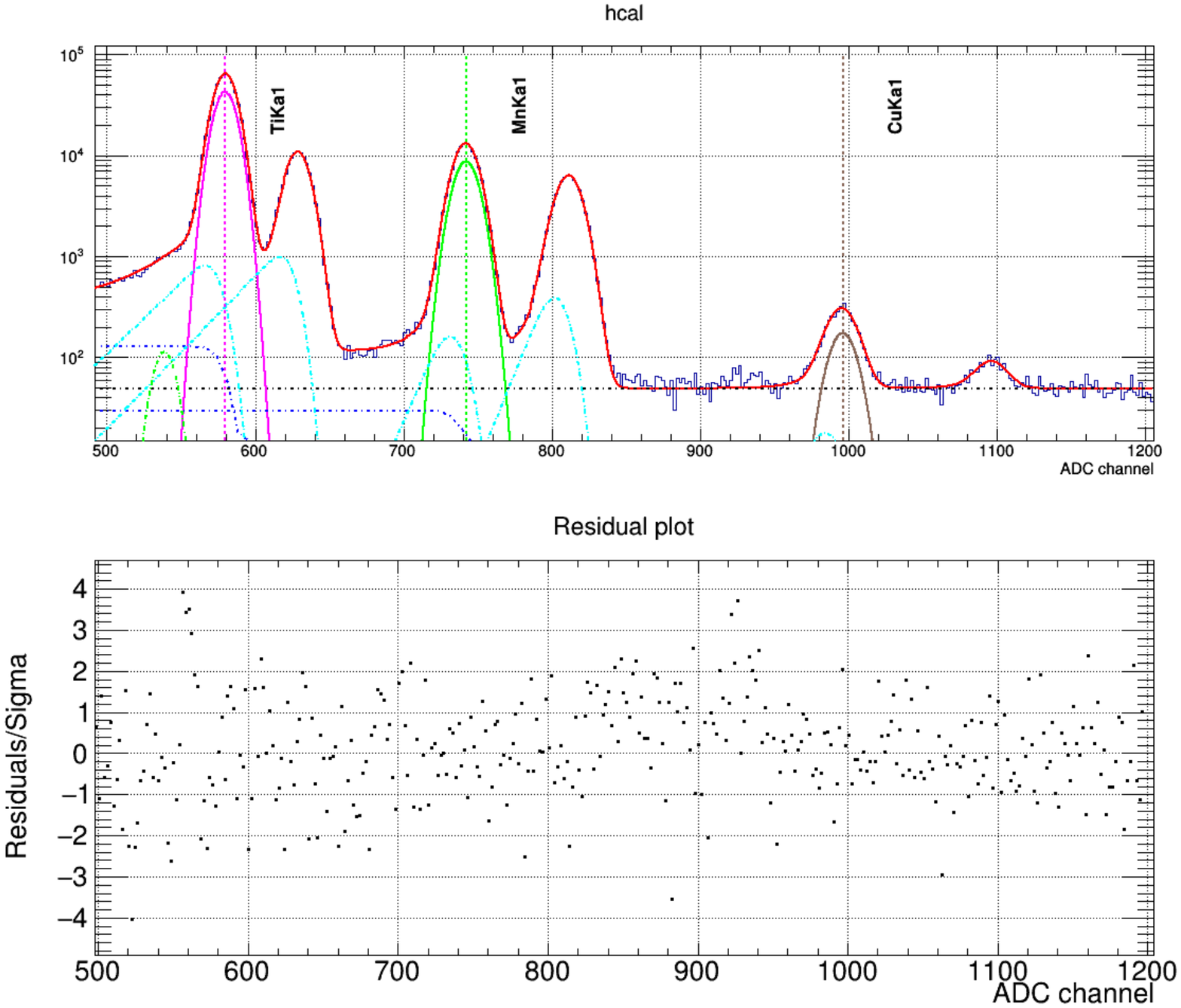}
  \caption{Energy spectrum of one SDD of the VIP2 detup recorded in one week of data taking. The energy resolution (FWHM) in the region of the copper transition is about 190 eV .}\label{spectrum}
\end{figure}

The progress of the VIP2 experiment has  been reported in \cite{Pichler2016, Shi2016, Marton2015, Curceanu2016a}. In 2016 we collected data in a time period of $\sim$70 without current and $\sim$40 days with 100 A current. In fig. \ref{spectrum} a tyüical x-ray spectrum of one SDD of the detector array is displayed.

With the analysis technique of Ramberg and Snow \cite{Ramberg1990} we obtain a preliminary upper limit for the probability that the PEP is violated for electrons in copper

\begin{equation}\label{Prel-result}
  \beta^{2}/2 \leq 2.4 \times 10^{-29}
\end{equation}

\section{Summary and Outlook} 

The experimental program for testing a possible PEP violation for electrons made great progress in 2016. The use of a new type of SDDs as X-ray detectors can further enhance the sensitivity by providing larger sensitive area. Furthermore, the cooling can be made more simple changing from liquid argon to Peltier cooling.
Concerning the reduction of the X-ray background we will install a passive shielding with Teflon, lead and copper. Given a running time of 3 years and alternating measurement with and without current we expect to lower the upper limit of PEP violation by about two orders of magnitude.

\ack
We thank H. Schneider, L. Stohwasser, and D. Pristauz-Telsnigg from Stefan-Meyer-Institut for their
fundamental contribution in designing and building the VIP2 setup.
We acknowledge the very important assistance of the INFN-LNGS laboratory staff during all
phases of preparation, installation and data taking.
The support from the EU COST Action CA 15220 is gratefully acknowledged.
We thank the Austrian Science Foundation (FWF) which supports the VIP2 project with the
grants P25529-N20 and  W1252-N27 (doctoral college particles and interactions) and Centro Fermi for the grant “Problemi aperti nella meccania quantistica”.
Furthermore, this paper was made possible through the support of a grant from the
Foundational Questions Institute, FOXi (“Events” as we see them: experimental test of the
collapse models as a solution of the measurement problem) and a grant from the John Templeton
Foundation (ID 58158). The opinions expressed in this publication are those of the authors
and do not necessarily reflect the views of the John Templeton Foundation.

\section{References} 
\bibliographystyle{iopart-num} 
%
%\clearpage
%\textbf{\Large References}\\ \\

%
\bibliography{Bibliography.bib}{}

\providecommand{\newblock}{}
\begin{thebibliography}{10}
\expandafter\ifx\csname url\endcsname\relax
  \def\url#1{{\tt #1}}\fi
\expandafter\ifx\csname urlprefix\endcsname\relax\def\urlprefix{URL }\fi
\providecommand{\eprint}[2][]{\url{#2}}

\bibitem{Pauli1925}
Pauli W 1925 {\em Zeitschrift f{\"{u}}r Physik\/} {\bf 31} 765--783 ISSN
  0044-3328 \urlprefix\url{http://link.springer.com/10.1007/BF02980631}

\bibitem{Pauli1940}
Pauli W 1940 {\em Phys. Rev.\/} {\bf 58} 716--722
  \urlprefix\url{http://link.aps.org/doi/10.1103/PhysRev.58.716}

\bibitem{Luders1958}
L{\"{u}}ders G and Zumino B 1958 {\em Phys. Rev.\/} {\bf 110} 1450--1453
  \urlprefix\url{http://link.aps.org/doi/10.1103/PhysRev.110.1450}

\bibitem{Jackson2008a}
Jackson M~G 2008 {\em Physical Review D\/} {\bf 78} 126009 (\textit{Preprint}
  \eprint{0809.2733})
  \urlprefix\url{http://dx.doi.org/10.1103/PhysRevD.78.126009}

\bibitem{Balachandran2010}
Balachandran A~P, Joseph A and Padmanabhan P 2010 {\em Physical Review
  Letters\/} {\bf 105} 051601 ISSN 0031-9007
  \urlprefix\url{http://link.aps.org/doi/10.1103/PhysRevLett.105.051601}

\bibitem{Ramberg1990}
Ramberg E and Snow G~A 1990 {\em Physics Letters B\/} {\bf 238} 438--441 ISSN
  0370-2693
  \urlprefix\url{http://www.sciencedirect.com/science/article/pii/037026939091762Z}

\bibitem{Sperandio2008}
Sperandio L 2008 {\em {New experimental limit on the Pauli Exclusion Principle
  violation by electrons from the VIP experiment}\/} Ph.D. thesis

\bibitem{DiMatteo2005}
Matteo S~D and Sperandio L 2005 {VIP Technical Note IR - 4} Tech. rep.

\bibitem{Collaboration2004}
Collaboration V 2004 http://www.lnf.infn.it/esperimenti/vip
  \urlprefix\url{http://www.lnf.infn.it/esperimenti/vip}

\bibitem{Bartalucci2006}
Bartalucci S, Bertolucci S, Bragadireanu M, Bucci C, Cargnelli M, Catitti M,
  Curceanu C, {Di Matteo} S, Egger J~H, Ferrari N, Fuhrmann H, Guaraldo C,
  Iliescu M, Ishiwatari T, Laubenstein M, Marton J, Milotti E, Pietreanu D,
  Ponta T, Sirghi D, Sirghi F, Sperandio L, Widmann E and Zmeskal J 2006 {\em
  Physics Letters B\/} {\bf 641} 18--22 ISSN 03702693
  \urlprefix\url{http://linkinghub.elsevier.com/retrieve/pii/S0370269306009385}

\bibitem{Egger1993}
Egger J~P, Chatellard D and Jeannet E 1993 {Progress in Soft X-Ray Detection:
  The Case of Exotic Hydrogen} {\em Muonic Atoms and Molecules\/} (Basel:
  Birkh{\"{a}}user Basel) pp 331--344

\bibitem{Beer2005}
Beer G, Bragadireanu A~M, Cargnelli M, Curceanu-Petrascu C, Egger J~P, Fuhrmann
  H, Guaraldo C, Iliescu M, Ishiwatari T, Itahashi K, Iwasaki M, Kienle P,
  Koike T, Lauss B, Lucherini V, Ludhova L, Marton J, Mulhauser F, Ponta T,
  Schaller L~A, Seki R, Sirghi D~L, Sirghi F, Zmeskal J and Zmeskal J 2005 {\em
  Physical Review Letters\/} {\bf 94} 212302 ISSN 0031-9007
  \urlprefix\url{http://link.aps.org/doi/10.1103/PhysRevLett.94.212302}

\bibitem{Ishiwatari2006a}
Ishiwatari T, Beer G, Bragadireanu A, Cargnelli M, {Curceanu (Petrascu)} C,
  Egger J~P, Fuhrmann H, Guaraldo C, Iliescu M, Itahashi K, Iwasaki M, Kienle
  P, Lauss B, Lucherini V, Ludhova L, Marton J, Mulhauser F, Ponta T, Schaller
  L, Sirghi D, Sirghi F, Strasser P and Zmeskal J 2006 {\em Nuclear Instruments
  and Methods in Physics Research Section A: Accelerators, Spectrometers,
  Detectors and Associated Equipment\/} {\bf 556} 509--515 ISSN 01689002

\bibitem{Pichler2016}
Pichler A, Bartalucci S, Bazzi M, Bertolucci S, Berucci C, Bragadireanu M,
  Cargnelli M, Clozza A, Curceanu C, {De Paolis} L, {Di Matteo} S, D'Uffizi A,
  Egger J~P, Guaraldo C, Iliescu M, Ishiwatari T, Laubenstein M, Marton J,
  Milotti E, Pietreanu D, Piscicchia K, Ponta T, Sbardella E, Scordo A, Shi H,
  Sirghi D, Sirghi F, Sperandio L, Vazquez-Doce O, Widmann E and Zmeskal J 2016
  {\em Journal of Physics: Conference Series\/} {\bf 718} 052030
  (\textit{Preprint} \eprint{1602.00898})
  \urlprefix\url{http://iopscience.iop.org/article/10.1088/1742-6596/718/5/052030/meta}

\bibitem{Shi2016}
Shi H, Bazzi M, Beer G, Bellotti G, Berucci C, Bragadireanu A, Bosnar D,
  Cargnelli M, Curceanu C, Butt A, D'Uffizi A, Fiorini C, Ghio F, Guaraldo C,
  Hayano R, Iliescu M, Ishiwatari T, Iwasaki M, {Levi Sandri} P, Marton J,
  Okada S, Pietreanu D, Piscicchia K, {Romero Vidal} A, Sbardella E, Scordo A,
  Sirghi D, Sirghi F, Tatsuno H, {Vazquez Doce} O, Widmann E and Zmeskal J 2016
  {\em EPJ Web of Conferences\/} {\bf 126} 04045 ISSN 2100-014X
  (\textit{Preprint} \eprint{1601.02236})
  \urlprefix\url{http://arxiv.org/abs/1601.02236
  http://dx.doi.org/10.1051/epjconf/201612604045
  http://www.epj-conferences.org/10.1051/epjconf/201612604045}

\bibitem{Marton2015}
Marton J, Bartalucci S, Bertolucci S, Berucci C, Bragadireanu M, Cargnelli M,
  Curceanu C, Clozza A, Matteo S, Egger J~P, Guaraldo C, Iliescu M, Ishiwatari
  T, Laubenstein M, Milotti E, Pichler A, Pietreanu D, Piscicchia K, Ponta T,
  Scordo A, Shi H, Sirghi D, Sirghi F, Sperandio L, Doce O, Widmann E and
  Zmeskal J 2015 {\em Journal of Physics: Conference Series\/} {\bf 631} ISSN
  17426596 17426588
  \urlprefix\url{http://iopscience.iop.org/article/10.1088/1742-6596/631/1/012070/pdf}

\bibitem{Curceanu2016a}
Curceanu C, Bertolucci S, Bassi A, Bazzi M, Bertolucci S, Berucci C,
  Bragadireanu A, Cargnelli M, Clozza A, {De Paolis} L, {Di Matteo} S, Donadi
  S, D'uffizi A, Egger J~P, Guaraldo C, Iliescu M, Laubenstein M, Marton J,
  Milotti E, Pichler A, Pietreanu D, Piscicchia K, Ponta T, Scordo A, Shi H,
  Sirghi D, Sirghi F, Sperandio L, Doce O and Zmeskal J 2016 {\em International
  Journal of Quantum Information\/} {\bf 14} 1--10 ISSN 02197499
  \urlprefix\url{http://www.worldscientific.com/doi/abs/10.1142/S0219749916400177}

\end{thebibliography}

\end{document}